\documentclass[%
 reprint,
 superscriptaddress,
 amsmath,amssymb,
 aps,pra,
floatfix,
]{revtex4-2}
\usepackage{lipsum}
\usepackage{graphicx}
\usepackage{dcolumn}
\usepackage{bm}
\usepackage[colorlinks,linkcolor=blue,citecolor = blue,urlcolor=blue]{hyperref}
\usepackage{booktabs}
\usepackage{float}
\usepackage{csquotes}
\usepackage{abraces}
\usepackage{commath}
\usepackage{ulem}
\usepackage{braket}
\usepackage{color}
\usepackage{cleveref}
\begin{document}
\title{Generalized description of the spatio-temporal biphoton state in spontaneous
parametric down-conversion}

\author{Baghdasar Baghdasaryan}
\email{baghdasar.baghdasaryan@uni-jena.de}
\affiliation{Theoretisch-Physikalisches Institut, Friedrich Schiller University Jena, 07743 Jena, Germany}
\affiliation{Helmholtz-Institut Jena, 07743 Jena, Germany}

\author{Carlos Sevilla-Gutiérrez}
 \affiliation{Fraunhofer Institute for Applied Optics and Precision Engineering IOF, 07745 Jena, Germany}

\author{Fabian Steinlechner}
\email{fabian.steinlechner@uni-jena.de}
 \affiliation{Fraunhofer Institute for Applied Optics and Precision Engineering IOF, 07745 Jena, Germany}
 \affiliation{Abbe Center of Photonics, Friedrich Schiller University Jena, 07745 Jena, Germany}

\author{Stephan Fritzsche}
\affiliation{Theoretisch-Physikalisches Institut, Friedrich Schiller University Jena, 07743 Jena, Germany}
\affiliation{Helmholtz-Institut Jena, 07743 Jena, Germany}
 \affiliation{Abbe Center of Photonics, Friedrich Schiller University Jena, 07745 Jena, Germany}
\date{\today}

\begin{abstract}
Spontaneous parametric down-conversion (SPDC) is a widely used source for photonic entanglement. Years of focused research have led to a solid understanding of the process, but a cohesive analytical description of the paraxial biphoton state has yet to be achieved. We derive a general expression for the spatio-temporal biphoton state that applies universally across common experimental settings and correctly describes the nonseparability of spatial and spectral modes. We formulate a criterion on how to decrease the coupling of the spatial from the spectral degree of freedom by taking into account the Gouy phase of interacting beams. This work provides new insights into the role of the Gouy phase in SPDC, and also into the preparation of engineered entangled states for multidimensional quantum information processing.
\end{abstract}

\maketitle
\section{Introduction}\label{introduction}
Photon pairs generated via spontaneous parametric down-conversion (SPDC) have provided an experimental platform for fundamental quantum science \cite{doi:10.1063/5.0023103} and figure prominently in applications in quantum information processing, including recent milestone experiments in photonic quantum computing \cite{doi:10.1126/science.abe8770}.

Several works in recent years have addressed the challenge of tailoring the spectral and spatial properties of $signal$ and $idler$ photons generated via SPDC in theory and experiment. In the spatial domain, that is, the transverse momentum space, much of this work was motivated by the objective of improving fiber coupling efficiency \cite{PhysRevA.83.023810,srivastav2021characterising} or the dimensionality of spatial entanglement \cite{Krenn6243,PhysRevA.102.052412,PhysRevApplied.14.054069}. In the spectral domain, the motivation was usually to engineer pure spectral states, which are crucial for protocols based on multiphoton interference \cite{Caspani2017}. This has been performed either by tailoring the nonlinearity of the crystal \cite{Graffitti:18} or by using counterpropagating photon pair generation in periodically poled waveguides \cite{Luo:20}. The frequency degree of freedom (DOF) has also been used to generate entangled states via spatial shaping of the pump beam \cite{Francesconi2021} or by transferring polarization into color entanglement \cite{PhysRevLett.103.253601}. The spatial shaping of the pump beam has been also used in Hong-Ou-Mandel interference experiments, in order to control the two-photon interference behavior \cite{PhysRevLett.90.143601}.

Closed expressions for the state emitted by SPDC in bulk crystals have been derived using very special techniques and approximations, such as the narrowband \cite{PhysRevA.83.033816}, thin-crystal \cite{Yao_2011,PhysRevA.103.063508} or plane wave approximations \cite{PhysRevLett.99.243601}, where either the spectral or spatial biphoton state is considered. However, from the $X$-shaped spatio-temporal correlations \cite{PhysRevLett.102.223601,PhysRevLett.109.243901}, the spatial and spectral properties of SPDC have been known to be coupled. The $X$-shaped spatio-temporal correlation implies that if the twin photons are collected from different positions, they are detected with a certain time delay. In contrast, if the photons are detected at the same position, the time delay is very short (a few nanoseconds) \cite{Zhang2017}. 

To date, models that address both spectrum and space together have been limited to approximate phase matching functions \cite{Osorio_2008} or numerical calculations \cite{PhysRevA.86.053803}. The work \cite{PhysRevLett.102.223601} investigated the quite general phase matching function, but the pump beam was limited to monochromatic plane wave.

Here, we present a simple-to-use closed expression for the biphoton state. The approach describes the full spectral and spatial properties of all interacting beams and applies to a wide range of experimental settings. The analytical treatment of the biphoton state decomposed into discrete Laguerre Gaussian (LG) modes also provides a deeper insight into the role of the Guoy phase in PDC. Especially, the spectral response of spatial modes in SPDC is determined by the Gouy phase of the pump, signal and idler beams. We will also show that the  Gouy phase can be used to control the coupling strength of spatial and spectral DOF in parametric down-conversion (PDC). Next to providing an intuitive understanding, we also demonstrate the utility of the expression for quantum state engineering in spatial DOF for multidimensional quantum information processing. 

\section{Theoretical Methods}

Let us start with the basic expressions of SPDC. We can make use of the paraxial approximation, since typical optical apparatuses support only paraxial rays about a central axis. In the paraxial regime, the longitudinal and transverse components of the wave vector can be treated separately $\bm{k}=\bm{q}+k_z(\omega)\bm{z}$.
Consequently, the biphoton state in the momentum space can be represented by the following expression \cite{PhysRevA.62.043816,WALBORN201087,Karan_2020}
\begin{align}\label{SPDCstate}
    \ket{\Psi} = \iint & d\bm{q}_s \: d\bm{q}_i \:d\omega_s \: d\omega_i\: \Phi(\bm{q}_s,\bm{q}_i,\omega_s,\omega_i)\nonumber\\&
  \hat{a}^{\dagger}_s(\bm{q}_s,\omega_s)\:\hat{a}^{\dagger}_i(\bm{q}_i,\omega_i)\ket{vac}.
\end{align}
Equation \eqref{SPDCstate} refers to the generation of photon pairs with energies $\omega_{s,i}$ and transverse momenta $\bm{q}_{s,i}$ from the vacuum state $\ket{vac}$. The biphoton mode function $\Phi(\bm{q}_s,\bm{q}_i,\omega_s,\omega_i)$ contains the rich high-dimensional spatio-temporal structure of SPDC that arises from the coupling between the wave vectors of the pump, signal, and idler beams. 

\subsection{Biphoton state decomposed in Laguerre Gaussian basis}

The transverse spatial  \cite{WALBORN201087,PhysRevA.83.052325} and frequency DOF \cite{PhysRevA.105.052429} have been successfully used in continuous variable information processing. However, in practical experimental settings, the continuous variable space is more often discretized using a set of modes. The proper choice of a set reduces the number of dimensions needed to describe the state. Moreover, discrete modes are easy to manipulate and detect using efficient experimental techniques \cite{Bolduc2013,Eckstein2011}. Since the projection of the orbital angular momentum (OAM) is conserved in SPDC \cite{Mair2001}, it is convenient to decompose the biphoton state into LG modes $\ket{p,\ell,\omega}=\int  d\bm{q}\, \mathrm{LG}_{p}^{\ell}(\bm{q})\,  \hat{a}^{\dagger}(\bm{q},\omega) \ket{vac} $, which are eigenstates of OAM \cite{
doi:10.1126/science.1227193}:
\begin{align}\label{decomposition}
    \ket{\Psi}= &\iint  \:d\omega_s \: d\omega_i\: \nonumber \\& \sum_{p_s,p_i=0}^{\infty}\: \sum^{\infty}_{\ell_s,\ell_i=-\infty}C_{p_s,p_i}^{\ell_s,\ell_i} \ket{p_s,\ell_s,\omega_s}\ket{p_i,\ell_i,\omega_i},
\end{align}
where the coincidence amplitudes are calculated from the overlap integral $ C^{\ell_s,\ell_i}_{p_s,p_i} = \braket{p_s,\ell_s,\omega_s;p_i,\ell_i,\omega_i |\Psi}$,
\begin{align}\label{coe1}
    C^{\ell_s,\ell_i}_{p_s,p_i} 
    =    \iint  d\bm{q}_s \: d\bm{q}_i \: \Phi(\bm{q}_s,\bm{q}_i,\omega_s,\omega_i)\:[\mathrm{LG}_{p_s}^{\ell_s}(\bm{q}_s)]^*\nonumber&\\
        \times  [\mathrm{LG}_{p_i}^{\ell_i}(\bm{q}_i)]^*.
\end{align}
The angular distribution of an LG mode in the momentum space is given by
\begin{align} \label{LG}
  \mathrm{LG}_{p}^{\ell}(\rho,\varphi)
    =&e^{\frac{-\rho^2\,w^2}{4}}\,e^{i\ell\,\varphi}\,\sum_{u=0}^p\, T_u^{p,\ell}\, \rho^{2k+\abs{\ell}}
\end{align} 
with $ T_u^{p,\ell}$ being
\begin{align*}
   T_u^{p,\ell}= &\sqrt{\frac{p!\,(p+|\ell|)!}{\pi}}\,
   \biggr(\frac{ w}{\sqrt{2}}\biggl)^{2u+|\ell|+1}\,\frac{(-1)^{p+u}(i)^{\ell}}{(p-u)!\,(\abs{\ell}+u)!\,u!},
\end{align*}
and where $\rho$ and $\varphi$ stand for the cylindrical coordinates $\bm{q}=(\rho,\varphi)$. 
The summations in Eq. \eqref{decomposition} run over the LG mode numbers $p$ and $\ell$ associated with the radial momentum and the OAM projection, respectively. Except for the fact, that we now deal with summations instead of integrations, this discretization will also help us to understand the coupling of spatial and spectral DOF in the frame of the Gouy phase of LG modes. Note that we discretize only the transverse spatial DOF, but in principle, it is also possible to discretize the frequency DOF \cite{Gil-Lopez:21}.

The construction of the biphoton state reduces to the calculation of the coincidence amplitudes $C^{\ell_s,\ell_i}_{p_s,p_i}$, which in turn depend on the mode function $\Phi(\bm{q}_s,\bm{q}_i,\omega_s,\omega_i)$. A compact expression for the mode function can be derived if the experimental setup and geometry is fixed.

\subsection{Geometry and mode function}
Here, we consider the scenario when a coherent laser beam propagates along the $z$ axis and is focused in the middle of a nonlinear crystal placed at $z=0$. Signal and idler fields propagate close to the pump direction, known as the quasicollinear regime. The crystal and the pump beam have typical transverse cross sections in the order of millimeters and micrometers, respectively. Hence, we assume that the crystal compared to the pump beam is infinitely extended in the transverse direction, which enforces the conservation of the transverse momentum, $\bm{q}_p=\bm{q}_{s}+\bm{q}_{i}$ \cite{PhysRevA.62.043816}. Taking into account also the energy conservation $\omega_p=\omega_s+\omega_i$, the mode function can be written as \cite{Karan_2020}
\begin{align} \label{phasefunction}
 \Phi(\bm{q}_s,\bm{q}_i,\omega_s,\omega_i)=&N_0\,\mathrm{V}_p(\bm{q}_s+\bm{q}_i)\,\mathrm{S}_p(\omega_s+\omega_i)\nonumber\\&\times\int_{-L/2}^{L/2} dz\:\exp{\biggr[iz(k_{z,p}-k_{z,s}-k_{z,i})\biggl]},
  \end{align} 
where $N_0$ is the normalization constant, $\mathrm{V}_p(\bm{q}_p)$ is the spatial and $\mathrm{S}_p(\omega_p)$ the spectral distribution of the pump beam, and $L$ is the length of the nonlinear crystal along the $z$ axis.
The important component of the mode function \eqref{phasefunction} is the phase mismatch in the $z$ direction $\Delta k_z=k_{z,p}-k_{z,s}-k_{z,i}$, which characterizes the differences in the energies and momenta of the signal and idler photons. Therefore, careful calculation of $\Delta k_z$ is essential for the quantitative description of SPDC, which we will do next.

Experimentally generated lights are usually not monochromatic and contain a frequency distribution. Therefore, except for the central frequencies that meet energy conservation condition $\omega_{0,p}=\omega_{0,s}+\omega_{0,i}$, we expect a deviation from them, $\omega=\omega_0+\Omega$ with the assumption $\Omega\ll\omega_0$. Furthermore, in the paraxial approximation, the transverse component of the momentum is much smaller than the longitudinal component $|\bm{q}|\ll k$. Hence, we can apply the Taylor series on $k_z$ (Fresnel approximation) to $|\bm{q}|/k$ and also to small $\Omega$:
\begin{equation*}
    k_z=k(\Omega)\sqrt{1-\frac{|\bm{q}|^2}{k(\Omega)^2}}\approx k+\frac{\Omega}{u_g}+\frac{G\Omega^2}{2}-\frac{|\bm{q}|^2}{2k},
  \end{equation*}
where $u_g=1/(\partial k/\partial \Omega)$ is the group velocity and $G=\partial/\partial \Omega \,(1/u_g)$ is the group velocity dispersion, evaluated at the respective central frequency. Here, we also assume that the propagation is along a principal axis of the crystal, so we can ignore the Poynting vector walk-off of extraordinary beams in the crystal. Next,
we insert the corresponding $k_z$ of the pump, signal, and idler into the phase mismatch $\Delta k_z$ and arrive at
\begin{equation}\label{phaseMatching}
\Delta k_z= \Delta_{\Omega}+\rho_{s}^2\frac{k_p-k_s}{2k_pk_s}+\rho_{i}^2\frac{k_p-k_i}{2k_pk_i}
-\frac{\rho_{s}\rho_{i}}{k_p}\cos{(\varphi_i-\varphi_s)},
\end{equation}
where the frequency part $\Delta_{\Omega}$ is given by
\begin{align}\label{spectralStr}
\Delta_{\Omega}=&\frac{\Omega_s+\Omega_i}{u_{g,p}}-\frac{\Omega_s}{u_{g,s}}-\frac{\Omega_i}{u_{g,i}}+\frac{G_p(\Omega_s+\Omega_i)^2}{2}\nonumber\\&
-\frac{G_s\Omega_s^2}{2}-\frac{G_i\Omega_i^2}{2}.
\end{align}
We used in Eq. \eqref{phaseMatching} the relation $\rho_p^2=\rho^2_s+\rho^2_i+2\rho_s\rho_i\cos{(\varphi_i-\varphi_s)}$ and assumed momentum conservation for central frequencies $\Delta k=k_p-k_s-k_i=0$. The condition $\Delta k=0$ ensures constructive interference in the crystal between the pump, signal, and idler beams, which is usually performed with birefringent crystals \cite{Karan_2020} or more recently by periodic poling along the crystal axis, $k_p-k_s-k_i-2 \pi /\Lambda=0$, where $\Lambda$ is the poling period \cite{doi:10.1063/1.123408}.
The remaining components of the mode function \eqref{phasefunction} that we should still fix are the pump characteristics. We model the angular distribution of the pump with an LG beam. The advantage of this choice is that an arbitrary paraxial optical field can be expressed as a sum of LG beams $\sum_n a_n \mathrm{LG}_{p_n}^{\ell_n} $ with $\sum_n |a_n|^2=1$ by using their completeness relation. Thus, the theory developed for the LG pump can be easily extended to SPDC with a particular pump. The amplitudes $\eqref{expression}$ can then be upgraded to revised amplitudes $\sum_n a_n C_n$, which follows from Eq. \eqref{coe1}. Finally, the temporal distribution is modeled with a Gaussian envelope of pulse duration $t_0$, $\mathrm{S}_p(\omega_p)=\exp{[-(\omega_p-\omega_{0,p})^2\,t_0^2/4]}\, t_0/\sqrt{\pi}$ \cite{PhysRevA.56.1627}, but which can be extended to any arbitrary pump spectrum.
 
\subsection{Derivation of coincidence amplitudes} We can now substitute Eqs. \eqref{LG}-\eqref{spectralStr} into Eq. \eqref{coe1} and calculate the coincidence amplitudes:
\begin{widetext}
\begin{align} \label{expansionfirst}
 C_{p,p_s,p_i}^{\ell,\ell_s,\ell_i}=N_0\sum_{u=0}^p\,\sum_{s=0}^{p_s}\,\sum_{i=0}^{p_i}\,T_u^{p,\ell}\: (T_s^{p_s,\ell_s})^* \:(T_i^{p_i,\ell_i})^*\: \int\,dz\,d\rho_s\,d\rho_i\,d\varphi_s\,d\varphi_i\,
      \Theta(z,\rho_s,\rho_i,\varphi_i-\varphi_s)
   e^{i\ell\varphi_s}\,e^{i(-\ell_s\varphi_s-\ell_i\varphi_i)},
\end{align}
where we used a revised notation for the coincidence amplitudes $C_{p,p_s,p_i}^{\ell,\ell_s,\ell_i}$ to indicate the mode numbers of the pump. The function $\Theta(z,\rho_s,\rho_i,\varphi_i-\varphi_s)$ is defined as
\begin{align}
 \Theta(z,\rho_s,\rho_i,\varphi_i-\varphi_s)=&
   [\rho^2_s+\rho^2_i+2\rho_s\rho_i\cos{(\varphi_i-\varphi_s)}]^\frac{2u+(\abs{\ell}-\ell)}{2}\,\rho_s^{\,\abs{\ell_s}+2s+1}\,
    \rho_i^{\,\abs{\ell_i}+2i+1}\,(\rho_s+\rho_i\,e^{i(\varphi_i-\varphi_s)})^{\ell}
     \nonumber\\&\times
     \exp{\biggl[-\frac{[\rho^2_s+\rho^2_i+2\rho_s\rho_i\cos{(\varphi_i-\varphi_s)}]\,w^2}{4}-\frac{\rho_s^2\,w_s^2}{4}-\frac{\rho_i^2\,w_i^2}{4}\biggr]}\,\frac{t_0}{\sqrt{\pi}}e^{-\frac{t_0^2(\Omega_s+\Omega_i)^2}{4}} 
      \nonumber \\& \times
      \exp{\biggr[iz\biggr(\Delta_{\Omega}+\rho_s^2\frac{k_p-k_s}{2k_pk_s}+\rho_i^2\frac{k_p-k_i}{2k_pk_i}-\cos{(\varphi_i-\varphi_s)}\frac{\rho_s\rho_i}{k_p}\biggl)\biggl]}\label{bigexpression}.    \end{align}
\end{widetext}
In Eq. \eqref{bigexpression}, the polar angle $\varphi$ of the pump beam has been expressed as a function of signal and idler coordinates,
\begin{equation*}
e^{i\,\ell\,\varphi}=(\cos{\varphi}+i\sin{\varphi})^{\ell}=\frac{e^{i\ell\varphi_s}}{\rho_p^{\ell}}\,(\rho_s+\rho_i\,e^{i(\varphi_i-\varphi_s)})^{\ell},   \end{equation*}
by taking into account the conservation of transverse momentum,
\begin{equation*}
\bm{q}_p=\bm{q}_{s}+\bm{q}_{i}=
\begin{pmatrix}
\rho_{s}\cos{\varphi_{s}}+\rho_{i}\cos{\varphi_{i}}\\
\rho_{s}\sin{\varphi_{s}}+\rho_{i}\sin{\varphi_{i}}
\end{pmatrix}.
\end{equation*}
The presentation of the coincidence amplitudes $C_{p,p_s,p_i}^{\ell,\ell_s,\ell_i}$ in Eq. \eqref{expansionfirst} with the function $\Theta(z,\rho_s,\rho_i,\varphi_i-\varphi_s)$ follows the goal to show the OAM conservation in SDPC. To do so, we expand the function $\Theta(z,\rho_s,\rho_i,\varphi_i-\varphi_s)$ as superposition of plane waves with the phases $\exp{[i\ell^{'}(\varphi_i-\varphi_s)]} $(Fourier series with complex coefficients),
 \begin{equation}\label{phasedif}
      \Theta(z,\rho_s,\rho_i,\varphi_i-\varphi_s)=\sum_{\ell^{'}=-\infty}^{\infty}f_{\ell^{'}}(z,\rho_s,\rho_i)e^{i\ell^{'}(\varphi_i-\varphi_s)}.
 \end{equation}
We substitute expression \eqref{phasedif} into Eq. \eqref{expansionfirst} and perform the integration over the polar angles $\varphi_s$ and $\varphi_i$:
\begin{align}
    \sum^\infty_{\ell^{'}=-\infty}f_{\ell^{'}}(z,\rho_s,\rho_i)\int_0^{2\pi}\int_0^{2\pi}   e^{i\ell\varphi_s}\,e^{i(-\ell_s\varphi_s-\ell_i\varphi_i)}\nonumber\\
    \times e^{i\ell^{'}(\varphi_i-\varphi_s)}d\varphi_sd\varphi_i \propto \delta_{\ell^{'},\ell-\ell_s}\delta_{\ell^{'},\ell_i}.\label{delta}
\end{align}
As expected, the Kronecker delta functions appear in Eq. \eqref{delta} which enforce the conservation of OAM $\ell-\ell_s=\ell_i$. This conservation is not valid out of the quasicollinear regime \cite{MOLINATERRIZA2003155} because of the spin-orbital angular momentum coupling in the nonparaxial regime \cite{PhysRevA.99.023403}. In a non-collinear regime, the total angular momentum should remain conserved, which can be a future topic to study.

Going back to expression \eqref{expansionfirst}, we now calculate the integration over polar coordinates $\varphi_{s,i}$ explicitly. For simplicity, we consider the coincidence amplitudes $ C_{p,p_s,p_i}^{\ell,\ell_s,\ell_i}$ for positive OAM number of the pump beam $\ell \geq 0$. The coincidence amplitude for $\ell< 0$ is then given by $ C_{p,p_s,p_i}^{\ell,\ell_s,\ell_i}= (C_{p,p_s,p_i}^{-\ell,-\ell_s,-\ell_i})^*$, which follows from Eq. \eqref{coe1}. Furthermore, the two brackets on the first line in Eq. \eqref{bigexpression} should be rewritten as finite sums by using the binomial formula. For instance, the first bracket is written as
\begin{align*}
    [\rho^2_s+\rho^2_i+2\rho_s\rho_i\cos{(\varphi_i-\varphi_s)}]^u=\sum_{m=0}^{u}
  \binom{u}{m}(\rho^2_s+\rho^2_i)^{u-m}&\\
  \times [2\rho_s\rho_i\cos{(\varphi_i-\varphi_s)}]^m.
\end{align*}

The \textit{cosine} function can be expressed as the sum of two exponential functions by using Euler's formula, which should be again expressed as a Binomial sum. After this step, the angular integration takes the form of the integral representation of the Bessel function of the first kind \cite{YOUSIF1997199}
\begin{equation*}
\frac{1}{2\pi}\int^{2\pi}_0 e^{i n\varphi \pm iz\cos{(\varphi-\varphi^{\prime})}
    }d\varphi=   (\pm i)^n e^{i n\varphi^{\prime}} J_n(z).
\end{equation*}
Next, the sum representation of the Bessel function should be used
\begin{equation}
    J_n(z)= \sum_{k=0}^{\infty} \frac{(-1)^k}{k!\, \Gamma (k+n+1)}\biggl(\frac{z}{2}\biggr)^{2k+n},\label{Bsum}
\end{equation}
which transforms the integration over the radial coordinates into %
\begin{equation*}
    \int_0^{\infty} d\rho\,\rho^n e^{-a \,\rho^2}=\frac{\Gamma(\frac{n+1}{2})}{2a^{\frac{n+1}{2}}}.
\end{equation*}
The final result is achieved via summing over $k$ from Eq. \eqref{Bsum} by using the definition of the \textit{Regularized} hypergeometric function \cite{Hypergeometric2F1}. The coincidence amplitudes read for $\ell \geq 0$ as 
\begin{align} \label{expression}
    C_{p,p_s,p_i}^{\ell,\ell_s,\ell_i}
     = & N_0\,\pi^{3/2}\:t_0\: e^{-\frac{t_0^2(\Omega_s+\Omega_i)^2}{4}} \: \delta_{\ell,\ell_s+\ell_i} \nonumber\\
   &   \sum_{u=0}^{p}\sum_{s=0}^{p_s}\sum_{i=0}^{p_i} T_u^{p,\ell}\: (T_s^{p_s,\ell_s})^* \:(T_i^{p_i,\ell_i})^*\: \sum_{n=0}^{\ell}\sum_{m=0}^{u}\nonumber\\
   &  
  \binom{\ell}{n}\binom{u}{m}\sum_{f=0}^{u-m}\sum_{v=0}^{m}\: \binom{u-m}{f}\binom{m}{v} \Gamma[h]\:\Gamma[b]\nonumber\\
   &  
  \int_{-L/2}^{L/2}dz\:e^{i z\, \Delta_{\Omega}}\:\frac{D^{d}}{H^{h}\: B^{b}}\: {_2}{\Tilde{F}}_1\biggl[h,b, 1+d,\frac{D^2}{H \,B
  }\biggl]
\end{align} 
and $ C_{p,p_s,p_i}^{\ell,\ell_s,\ell_i}= (C_{p,p_s,p_i}^{-\ell,-\ell_s,-\ell_i})^*$ for $\ell < 0$. The function ${_2}{\Tilde{F}}_1$ is known as the \textit{regularized} \textit{hypergeometric} function \cite{Hypergeometric2F1}. The missing coefficients of Eq. \eqref{expression} are given by
\begin{eqnarray*}
 H &=& \frac{w_p^2}{4}+\frac{w_s^2}{4}-i z\frac{k_p-k_s}{2k_p k_s}, \qquad  D = -\frac{ w_p^2}{4}-iz\frac{1}{2k_p},               \\[0.1cm]
  B& = &  \frac{w_p^2}{4}+\frac{w_i^2}{4}-i z\frac{k_p-k_i}{2k_p k_i}, \qquad d =\ell_i+m-n-2v,  \\[0.1cm]
  h & = & \frac{1}{2}(2+2s+\ell+\ell_i+2(-f+u)-2n-2v+\abs{\ell_s}), \\[0.1cm]
  b &=& \frac{1}{2}(2+2f+2i+\ell_i+2m-2v+\abs{\ell_i}),
\end{eqnarray*}
where $w_p$, $w_s$ and $w_i$ are the beam waists of the pump signal and the idler beams, respectively. 

Expression \eqref{expression} for the coincidence amplitudes as a function of the pump mode constitutes the main result of this work. It allows the spatial and spectral emission profiles to be reconstructed mode by mode and is applicable in any experimental setting that exhibits cylindrical symmetry. It can be readily used to calculate many characteristics of SPDC: joint spectral density, photon bandwidths, pair-collection probability, heralding ratio, spectral and spatial correlation, etc. Previously, these could only be achieved through numerical calculations or for special cases with a limited scope of applicability. The experimental demonstration of Eq. \eqref{expression} has already been presented in Ref. \cite{carlos}, where we also showed how the coupling of spatial and spectral DOF deteriorates the spatial entanglement but can be compensated directly by a proper choice of the collection mode.
\subsection{Gouy phase and spatio-temporal coupling}
The spatio-temporal coupling encoded in Eq. \eqref{expression} is a fundamental feature of SPDC. However, the usual applications in quantum optics utilize either the spatial or spectral DOF, neglecting the correlation between them. Nevertheless, this coupling remains a fundamental issue in many protocols based on entangled photon sources, where any distinguishability arising
from not-considered DOF reduces the coherence of the state. Next, we will illustrate the utility of expression \eqref{expression} in the frame of possible decoupling of spatial and spectral DOF $\Phi(\bm{q}_s,\bm{q}_i,\omega_s,\omega_i)= \Phi_{\bm{q}}(\bm{q}_s,\bm{q}_i)\Phi_{\omega}(\omega_s,\omega_i)$. We will show that this decoupling is closely related to the Gouy phase of interacting beams.

The role of the Gouy phase in nonlinear processes has been investigated before. For instance, in SPDC, the change of the Gouy phase $\psi_G(z)=(N+1)\arctan(z/z_R)$ within the propagation distance has been used to control the relative phase of two different LG modes of measurement basis \cite{PhysRevLett.101.050501,DEBRITO2021126989}. Here, $N$ is the combined LG mode number $N=2p+\abs{\ell}$ and $z_R$ is the Rayleigh length. In four-wave mixing (FWM), the conversion behavior between LG modes is strongly affected by the Gouy phase \cite{PhysRevA.103.L021502}. The authors observed that the existence of a relative Gouy phase between modes with different mode numbers $N$ leads to a reduced FWM efficiency. 

Here, we have a similar situation: pump, signal, and idler fields acquire different Gouy phases along with propagation in the crystal due to different mode numbers $N$, causing a reduced efficiency of mode down-conversion.
We expect intuitively that the shape of the spectrum of spatial modes is affected by the relative Gouy phase of interacting beams. This is still a guess and requires proof.
We consider for simplicity the scenario in which the Rayleigh lengths of the three beams are equal $z_{R,p}=z_{R,i}=z_{R,s}$ and fixed. This condition matches the Gouy angle $\arctan(z/z_R)$ for all beams. Hence, the relative Gouy phase can be written as
\begin{equation*}
    \psi_{G,p}-\psi_{G,s}-\psi_{G,i}=(N_p-N_s-N_i-1)\arctan(z/z_R).
\end{equation*}
This implies that the Gouy phase is fully defined by the relative mode number $N_R= N_p-N_s-N_i$. If the Gouy phase is responsible for different spectral dependencies of the coincidence amplitudes $C_{p,p_s,p_i}^{\ell,\ell_s,\ell_i}(\Omega_s$,$\Omega_i)$, the shape of the spectrum should remain the same for fixed $N_R$. Assuming $k_p= 2 k_s$, Eq. \eqref{expression} transforms into

\begin{equation}
 C_{p,p_s,p_i}^{\ell,\ell_s,\ell_i}(\Omega_s,\Omega_i)\propto   \int_{-L/2}^{L/2}dz\:e^{i z\, \Delta_{\Omega}}\:\frac{(i2z+k_p w_p^2)^{N_R}}{(-i2z+k_p w_p^2)^{N_R+1}}. \label{gouy}
\end{equation}
We see from Eq. \eqref{gouy} that the spectral response of $C_{p,p_s,p_i}^{\ell,\ell_s,\ell_i}(\Omega_s,\Omega_i)$ encoded only in the term $e^{i z\, \Delta_{\Omega}}$ remains unaffected up to a constant if $N_R$ is fixed. On the other hand, $N_R$ can be rewritten as 
\begin{equation}\label{Rnumber}
    N_R=\frac{ \psi_{G,R}}{\arctan(z/z_R)}+1,
\end{equation}
where $\psi_{G,R}$ is the relative Gouy phase $\psi_{G,p}-\psi_{G,s}-\psi_{G,i}$.   Therefore, it follows from Eqs. \eqref{gouy} and \eqref{Rnumber} that the spectral response of coincidence amplitudes is determined by the relative Gouy phase $\psi_{G,R}$ if the pump characteristics $z_R$, $w_p$, and $k_p$ are fixed. This is what we wanted to prove. Note that the simple form of Eq. \eqref{gouy} is due to the assumptions $k_p= 2 k_s$ and $z_{R,p}=z_{R,i}=z_{R,s}$. The analytical proof for the general case requires more effort, which we omit here.

This proof brings us a step forward in the decoupling problem of spatial and spectral DOF: the decoupling can be achieved for a selected subspace of modes that possess the same relative Gouy phase. So, if a state is engineered that consists of modes with $N_R=const.$ assuming $z_{R,p}=z_{R,i}=z_{R,s}$, then the modes contributing to the state have the same spectrum, i.e., the state is separable. The question of decoupling of spatial and spectral DOF can be now reformulated: How do we engineer a state only consisting of modes with the same relative mode number $N_R$.
\begin{figure*}
\centering
\includegraphics[width=0.98\textwidth]{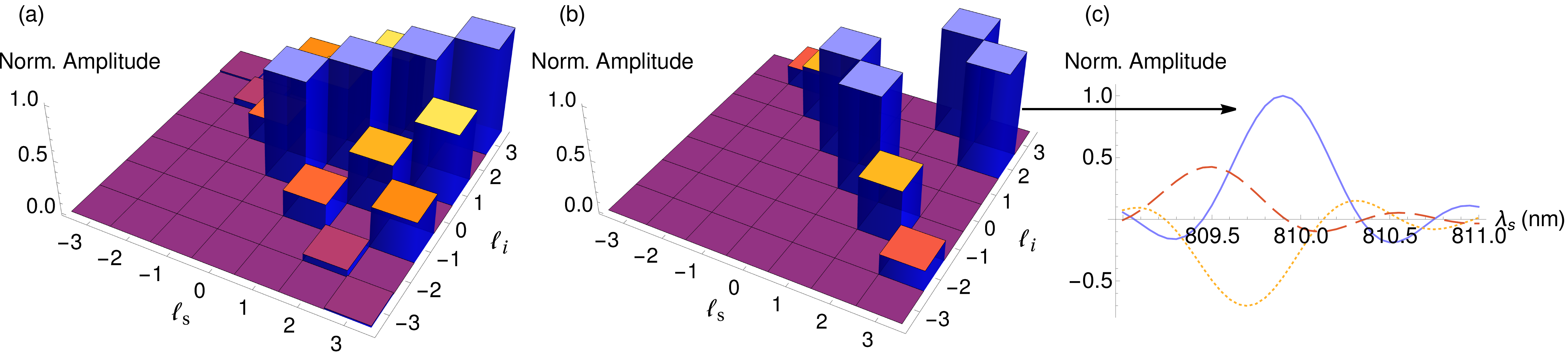}
\caption{Normalized high-dimensional entangled states (a) $\ket{\Psi^{\prime}_4}=\frac{1}{2}(\ket{0,0}+\ket{1,1}+\ket{2,2}+\ket{3,3})$ and (b) $\ket{\Psi_4}=\frac{1}{2}(\ket{0,1}+\ket{1,0}+\ket{2,3}+\ket{3,2})$. The state $\ket{\Psi_4}$ is maximally entangled in the subspace $\ell_s,\ell_i=0,1,2,3$, but not $\ket{\Psi^{\prime}_4}$ because of diagonal elements in the subspace. The state $\ket{\Psi^{\prime}_4}$ is maximally entangled in a smaller subspace, namely in the subspace consisting of only four modes $\{\ket{0,0},\ket{1,1},\ket{2,2},\ket{3,3}\}$. (c) The modes of signal (idler) involved in the state $\ket{\Psi_4}$  have the same spectrum (blue solid curve) compared to the modes out of the subspace shown in the same color as corresponding bars (dotted curve $\ket{2,-1}$ and dashed curve $\ket{3,-2}$). These curves correspond only to the spatial modes from (b).}\label{fig1.pdf}
\end{figure*}

\section{Engineering high-dimensional entangled states in OAM basis}
The state engineering in spatial DOF has been investigated theoretically in the thin crystal regime \cite{PhysRevA.67.052313,https://doi.org/10.1002/qute.202100066} and also implemented experimentally \cite{PhysRevA.98.060301,PhysRevA.98.062316}. In particular, three-, four-, and five-dimensional entangled states in OAM basis have been generated in Ref. \cite{PhysRevA.98.060301} using a superposition of LG beams for the pump. The correct superposition for the pump has been determined with a simultaneous perturbation stochastic approximation algorithm \cite{705889}.

We show in this section how to calculate the correct superposition of LG beams with Eq. \eqref{expression}, in order to generate entangled states in OAM basis including the states from Ref. \cite{PhysRevA.98.060301}. Our method is very straightforward and requires no optimization algorithm. In comparison to Refs. \cite{PhysRevA.67.052313,https://doi.org/10.1002/qute.202100066}, our results can be directly implemented in a real experiment, since we do not consider the thin crystal approximation. State engineering in the thin crystal regime is inefficient due to an infinite amount of spatial modes generated in the down-conversion.

\subsection{Determination of pump beam}
We consider the four-dimensional subspace $\ell_s,\ell_i=0,1,2,3$ and  $p_s=p_i=0$, which we refer to as $S_4$, with associated notation $\ket{p_s=0,\ell_s,\omega_s}\ket{p_i=0,\ell_i,\omega_i}:=\ket{\ell_s(\omega_s),\ell_i(\omega_i)}$. The goal is to engineer a four-dimensional maximally entangled state in this subspace. We model the pump beam as a superposition of LG beams,
\begin{eqnarray*}
   \mathrm{V}_p & = & \sum_{\ell}a_{\ell}\;\mathrm{LG}_{0}^{\ell},
\end{eqnarray*}
where the range of summation is determined with the possible minimal and maximal OAM values in the subspace, $\ell=[\mathrm{min}(\ell_s+\ell_i),\mathrm{max}(\ell_s+\ell_i)]$. The correct choice of the expansion amplitudes $a_{\ell}$ is now our task. Since the pump function appears in Eq. \eqref{decomposition}  linearly, the corresponding state in $S_4$ is given by
\begin{align*}
    \ket{\Psi_4}
     =  \sum^6_{\ell=0}a_{\ell} \:\sum^{3}_{\ell_s,\ell_i=0}C_{0,0,0}^{\ell,\ell_s,\ell_i} \ket{\ell_s,\ell_i}.
\end{align*}
The matrix representation of the state $\ket{\Psi_4}$ can clarify the right choice of the coefficients $a_{\ell}$. The matrix consists of $16$ elements and is given by the left-hand side of the following expression:
\begin{align} 
\begin{pmatrix}
\textcolor{blue}{a_0}\,C_{0,0} & \textcolor{red}{a_1}\,C_{1,0} &\textcolor{blue}{a_2}\,C_{2,0}& a_3\,C_{3,0}\\
\textcolor{red}{a_1\,}C_{0,1} & \textcolor{blue}{a_2}\,C_{1,1} &  a_3\,C_{2,1}& \textcolor{blue}{a_4}\,C_{3,1}\\
\textcolor{blue}{a_2}\,C_{0,2} & a_3\,C_{1,2} &  \textcolor{blue}{a_4}\,C_{2,2}&\textcolor{red}{a_5}\,C_{3,2}\\
a_3\,C_{0,3} & \textcolor{blue}{a_4}\,C_{1,3} &  \textcolor{red}{a_5}\,C_{2,3}& \textcolor{blue}{a_6}\,C_{3,3}
\end{pmatrix}\rightarrow
\begin{pmatrix}
0 & \textcolor{red}{1} &0&0\\
\textcolor{red}{1} & 0 &  0& 0\\
0& 0 &  0&\textcolor{red}{1}\\
0& 0 &  \textcolor{red}{1}& 0
\end{pmatrix},\label{matrix}
\end{align}
where we used the notation $ C_{0,0,0}^{\ell_s+\ell_i,\ell_s,\ell_i}= C_{\ell_i,\ell_s}$. The state becomes maximally entangled in this subspace if the matrix has exactly one entry of $1$ in each row and each column and $0$ elsewhere (permutation matrix). The right-hand side of expression \eqref{matrix} is such a state that can be engineered if we select $a_1=1/C_{0,1}\approx1/C_{1,0}$, $a_5=1/C_{2,3}\approx1/C_{3,2}$ and $a_0=a_2=a_3=a_4=a_6=0$, where we assumed degenerate SPDC $k_p\approx 2 k_s$. This choice leads to the state $\ket{\Psi_4}=\frac{1}{2}(\ket{0,1}+\ket{1,0}+\ket{2,3}+\ket{3,2})$. Thus, the state engineering is finished, where the coefficients of the pump superposition $\{a_1,a_5\}$ should be calculated with the expression \eqref{expression}. In the same way, the state $\ket{\Psi^{\prime}_4}=\frac{1}{2}(\ket{0,0}+\ket{1,1}+\ket{2,2}+\ket{3,3})$ from Ref. \cite{PhysRevA.98.060301} can also be engineered, if we select  $\{a_0,a_2,a_4,a_6\}$ to be equal to $\{1/C_{0,0},1/C_{1,1},1/C_{2,2},1/C_{3,3}\}$ and $a_1=a_3=a_5=0$.

The states $\ket{\Psi^{\prime}_4}$ and $\ket{\Psi_4}$ are presented in Figs. \ref{fig1.pdf}(a) and \ref{fig1.pdf}(b) with blue-colored bars on top. As we can see,  the modes contributing to the states $\ket{\Psi^{\prime}_4}$ and $\ket{\Psi_4}$ represent just a part of the full OAM emission (spiral bandwidth). Therefore, the postselection should be the final step in the engineering process, where undesirable modes are sorted out. Next, we should calculate the Schmidt number and the purity of the presented states, in order to evaluate the efficiency of the state preparation in the subspace $S_4$. We will use for all our calculations the same experimental parameters as in Ref. \cite{PhysRevA.98.060301}: $15$-$mm$-thick periodically poled
$\mathrm{KTiOPO}_4$ crystal designed for a collinear frequency
degenerate type-II phase matching, continuous-wave laser of wavelength $405$ \textit{nm} with beam waist $w_p=25$ $\mu m$ and detection modes of radius $w_{s,i}=33$ $\mu m$.
\subsection{Schmidt number and purity of subspace states}
We compare first the azimuthal Schmidt numbers of the states $\ket{\Psi_4}$ and $\ket{\Psi^{\prime}_4}$ in the subspace $S_4$. Obviously, the diagonal modes $\{\ket{0,2},\ket{2,0},\ket{1,3},\ket{3,1}\}$ in Fig. \ref{fig1.pdf} (a) are non-desirable and lead to a decrease of entanglement in $S_4$. Consequently, the state $\ket{\Psi^{\prime}_4}$ has an azimuthal Schmidt number less than $4$, $K=2.04$, while the Schmidt number of the state $\ket{\Psi_4}$ equals $4$. Therefore, the preparation of the state $\ket{\Psi_4}$ is more efficient in $S_4$ than for $\ket{\Psi^{\prime}_4}$. $K=4$ is necessary, but not a sufficient condition for a four-dimensional state to be maximally entangled. Additionally, the state should be pure. Hence, the state $\ket{\Psi_4}$ can be called maximally entangled in $S_4$, if it is also spatially pure.

In order to calculate the spatial purity of $\ket{\Psi_4}$, we need the reduced density matrix $\rho_{\bm{q}}$, which results from tracing over the frequency $\rho_{\bm{q}}=\mathrm{Tr}_{\Omega}(\rho)$. The fact of a continuous wave laser $\mathrm{S}_p(\omega_p)\propto\delta(\omega_p-\omega_{0,p})$ sets the condition $\Omega_s=-\Omega_i:=\Omega$, which transforms Eq. \eqref{decomposition} into
\begin{align}
    \ket{\Psi}= &\iint  \:d\Omega\: \sum^{\infty}_{\ell_s,\ell_i=-\infty}C_{\ell_s,\ell_i}(\Omega) \ket{\ell_s,\Omega}\ket{\ell_i,-\Omega}.
\end{align}
Now, we calculate the density matrix $\rho=|\Psi\rangle\langle\Psi|$ and then trace over the spectral domain, which yields:
\begin{equation}
\rho_{\bm{q}}=\sum_{\ell_s,\ell_i}\sum_{\Tilde{\ell_s},\Tilde{\ell_i}}A^{\Tilde{\ell_s},\Tilde{\ell_i}}_{\ell_s,\ell_i}|\ell_s,\ell_i\rangle\langle\Tilde{\ell_s},\Tilde{\ell_i}|,\label{density}
\end{equation}
where $A^{\Tilde{\ell_s},\Tilde{\ell_i}}_{\ell_s,\ell_i}=\int d\Omega\: C_{\ell_s,\ell_i}(\Omega)\:[C_{\Tilde{\ell_s},\Tilde{\ell_i}}(\Omega)]^*$ is the overlap integral of the spectra of the OAM modes. Equation \eqref{density} is very useful to calculate the spatial purity in small subspaces. 

We run summations in Eq. \eqref{density} over $\ell_s,\ell_i,\Tilde{\ell_s},\Tilde{\ell_i}=0,1,2,3$, renormalize the state, construct the density matrix of the subspace $\rho_{\bm{q},s}$ and calculate the purity $\mathrm{Tr}(\rho^2_{\bm{q},s})$. Here, the subscript $s$ indicates the consideration of the subspace $S_4$. In fact, the state $\ket{\Psi_4}$ is spatially pure, $\mathrm{Tr}(\rho^2_{\bm{q},s})=1$. The reason is very trivial: all modes that contribute to the state consist of only positive OAM numbers, which leads to the same $N_R=\abs{\ell}-\abs{\ell_s}-\abs{\ell_i}=0$ for all modes due to $\ell=\ell_s+\ell_i$. Moreover, the experimental parameters from Ref. \cite{PhysRevA.98.060301} satisfy the condition $z_{R,p}\approx z_{R,i}\approx z_{R,s}$. Hence, all modes have the same relative Gouy phase and consequently, the same spectrum, which is presented in Fig. \ref{fig1.pdf}(c) with the blue curve. This means that the spatial and spectral DOF are decoupled in $S_4$. Interestingly, even though the authors did not consider the spectral DOF, the prepared state from Ref. \cite{PhysRevA.98.060301} is also separable in space and frequency in the smaller subspace of only four modes $\{\ket{0,0},\ket{1,1},\ket{2,2},\ket{3,3}\}$. We suppose that the engineering of maximally entangled states in spatial DOF in a certain subspace enforces automatic decoupling in spatial and spectral DOF in that subspace.
\subsection{Purity and Schmidt number of the full biphoton state}
Obviously, the subspace $S_4$ is a part of the full SPDC emission. The first four OAM modes out of the subspace in Fig. \ref{fig1.pdf}(b), $\ket{2,-1}$ and $\ket{3,-2}$, possess different spectra in contrast to the modes in $S_4$, shown in Fig. \ref{fig1.pdf}(c) with dotted and dashed curves, respectively. The appearance of modes with distinguishable spectra indicates the inseparability of spatial and spectral DOF out of $S_4$. The more distinguishable modes contribute to the state, the stronger the spatio-temporal coupling. This, in turn, leads to reduced purity for the spatial biphoton state. Usually, narrowband filters are used in front of detectors to increase the purity of the spatial state. On one hand, the spectral filters improve the purity of the spatial state; on the other hand, they reduce the rate of entangled photons.

\begin{figure}[b]
\includegraphics[width=.47\textwidth]{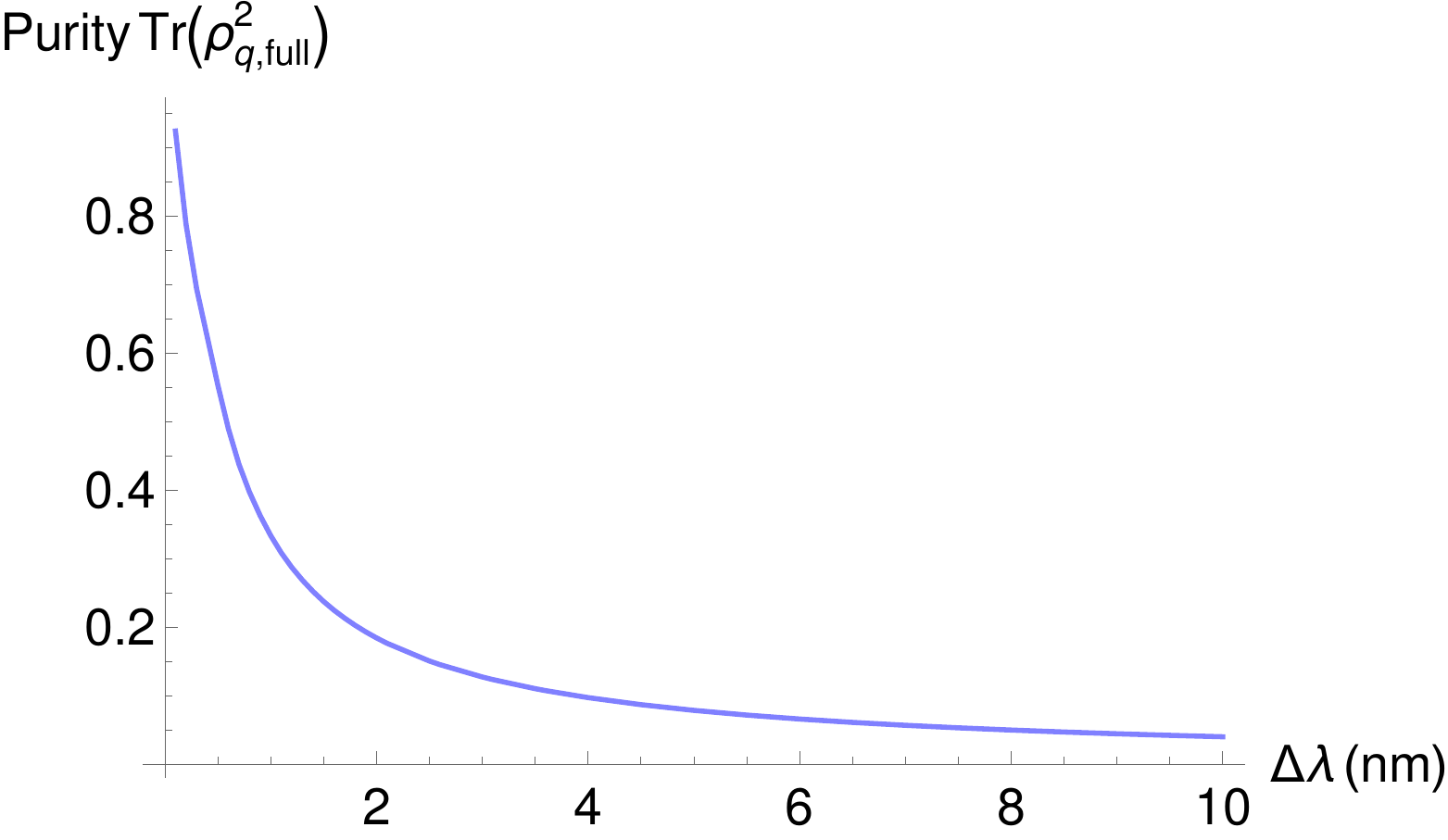}
\caption{Purity of the
spatial biphoton state depending on the bandwidth of spectral filter.}\label{fig2}
\end{figure}

We calculated the spatial purity $\mathrm{Tr}(\rho^2_{\bm{q},\mathrm{full}})$ \cite{Osorio_2008} of the full biphoton state \eqref{SPDCstate} depending on the filter bandwidth $\Delta \lambda$, in order to quantify the influence of spectral filters on the biphoton state. We chose as a pump the same beam, which leads to the state $\ket{\Psi_4}$. Very narrow filters are required to end up with a more or less pure state, as we can see from Fig. \ref{fig2}. For instance, a typical spectral filter with a bandwidth of $1$ \textit{nm} would leave the state in a mixed state of purity $0.33$.

The Schmidt number of the full spatio-temporal biphoton state is also different in comparison to the subspace state. The total Schmidt number can be calculated from the reduced density matrix in space and frequency for the signal by tracing over the idler $\rho_{\mathrm{signal}}=\mathrm{Tr}_{\mathrm{idler}}(\rho)$\cite{Osorio_2008}. The Schmidt number is then given by $K=1/\mathrm{Tr}(\rho^2_{\mathrm{signal}})$ \cite{computing}. The number of both spatial and spectral Schmidt modes in the range of frequencies $810 \pm 10$ \textit{nm} equals $140$, where $810$ \textit{nm} is the central frequency for signal and idler photons. In comparison, the number of Schmidt modes generated only at central frequency $810$ $nm$ equals $5.8$.

Finally, a small remark about the thin crystal regime: The spatio-temporal coupling is absent in the thin crystal regime $L \ll z_{R,p}$, since the biphoton state is independent of the crystal features. The problem with this regime is that it gives rise to a huge amount of spatial modes. Assume we keep all parameters the same as in Ref. \cite{PhysRevA.98.060301}, but change the crystal length to $L=1$ $\mu m$. The thin crystal regime is then well achieved according to Ref. \cite{PhysRevA.103.063508}.  The state becomes spatially pure, but possesses a large amount of Schmidt modes, $10^7$.
\section{CONCLUSION}
In summary, we derived a closed analytical expression for the biphoton spatio-temporal state in terms of the LG mode amplitudes. The expression readily reveals the dependence of the modal decomposition on frequency and thus correctly describes spectral-spatial coupling, a quintessential feature of SPDC. The expression provides a new understanding of how the Gouy phase is related to the decoupling of spatial and spectral DOF: the relative Gouy phase of the interacting beams fully defines the shape of the spectrum of down-converted photons. 

Engineering the modal decomposition of the pump beam can be used to engineer a high-dimensional OAM entanglement. State engineering can also be used to decrease the coupling between the spatial and spectral DOF, leading to an increase of the correlation stored in the spatial DOF. 
We thus hope that it will aid experimenters in the design and quantitative modeling of challenging experiments based on PDC.  

The authors thank Egor Kovlakov and Darvin Wanisch for very helpful discussions.
\bibliographystyle{apsrev4-1}
\bibliography{main.bib}      
\end{document}